\documentclass[aps,prl,nolongbibliography,twocolumn,floatfix,superscriptaddress]{revtex4-1}

\usepackage{amsmath}
\usepackage{amssymb}
\usepackage{amsthm}
\usepackage{mathrsfs}
\usepackage{mathtools}
\usepackage{graphicx}
\usepackage{float}
\usepackage{xcolor}
\graphicspath{{figures/}} 

\begin{document}

\title{Spreading processes on heterogeneous active systems: \\
spreading threshold, immunization strategies, and vaccination noise}


\date{\today}

\author{Benjam{\'i}n Marcolongo}
\affiliation{IFEG - CONICET and FaMAF - Universidad Nacional de C\'ordoba, C\'ordoba, Argentina.}
\author{Gustavo J. Sibona}
\affiliation{IFEG - CONICET and FaMAF - Universidad Nacional de C\'ordoba, C\'ordoba, Argentina.}
\author{Fernando Peruani}
\affiliation{Laboratoire de Physique Th{\'e}orique et Mod{\'e}lisation, UMR 8089, CY Cergy Paris Universit{\'e}, 95302 Cergy-Pontoise, France.}

\begin{abstract}
We study  spreading processes in two-dimensional systems of heterogeneous active agents that exhibit different individual active speeds. We obtain, combining kinetic and complex network theory, an analytical expression for the spreading threshold that depends not only on the first but also second moment of the speed distribution. Moreover, we prove that spreading can even occur for vanishing average active speed. Furthermore, we find that random vaccination strategies are ineffective in heterogeneous active systems, whereas targeted ones are effective. We also show that vaccination acts as (quenched) noise: it can decrease or increase the outbreak size. Our results offer  insights into how information propagates in heterogeneous populations of active agents.
\end{abstract}



\maketitle


Knowledge on how an internal variable -- e.g. an opinion, a behavioral state, or an epidemic state -- propagates in a system of mobile agents 
is of fundamental importance to understand not only disease spreading, but also information propagation in active systems~\cite{vicsek2012, marchetti2013}.  
Examples of contagion-like dynamics in active systems  -- where either an angle or a behavioral state is exchanged --  range from 
 synchronization of moving rotators~\cite{frasca08, riedel07, peruani2010mobility},   collective motion of animal groups~\cite{katz2011inferring, lopez2012behavioural, gomez2022intermittent}, to  
coordinated motion of human crowds~\cite{bain2019dynamic}. 
The empirical relevance of studying spreading processes in moving agent systems became particularly evident during COVID-19 pandemic 
with the need of forecasting, for policy making, the impact of lockdowns and social distancing on the spreading of the disease. 
Active Brownian particle (ABP) models have been used to evaluate the role of short-range human motion in epidemic spreading~\cite{peruani2008dynamics, peruani2019reaction, norambuena2020understanding, rodriguez2022epidemic, gu2024influence}, including the impact of social distancing~\cite{sajjadi2021social, te2020effects}, spreading across complex landscapes~\cite{forgacs2022using, zhu2023spatial}, role of agent velocity~\cite{rodriguez2019particle}, and contagion dynamics among collectively moving agents~\cite{zhao2022contagion}. 
Importantly, all these models consider (homogeneous) systems of identical active particles, where all individuals exhibit the same active speed. 
However, humans, animals, and biological systems in general are highly heterogeneous and individuals exhibit large inter-individual variability. 
For instance, in populations of locusts~\cite{knebel2019intra} and bacteria~\cite{otte2021statistics}, it has been reported  
large inter-individual variability of motility parameters, including large variation of individual active speeds, while in human populations it has been found 
a wide range of distances travelled by individuals within a fixed time window~\cite{gonzalez2008understanding}.  
Despite the evident fact that biological, real-world active systems are highly heterogeneous, the vast majority of 
 active matter studies~\cite{vicsek2012, marchetti2013}, including those that investigate the spreading of a disease  
 among active agents~\cite{peruani2008dynamics, peruani2019reaction, norambuena2020understanding, rodriguez2022epidemic, gu2024influence, sajjadi2021social, forgacs2022using, zhu2023spatial, rodriguez2019particle, zhao2022contagion}, and those where the disease is coupled to the agent activity~\cite{paoluzzi2020information, levis2020flocking, paoluzzi2018fractal, gascuel2023generic} consider populations of identical agents. 
Few exceptions are those that include particle-size polydispersity~\cite{huang2016epidemic, marchetti2016minimal, kumar2021effect, paoluzzi2022motility} and (static) social networks~\cite{miguel2018effects}. 
In summary, the physics of heterogeneous active agents -- despite its empirical relevance -- remains largely unexplored. 

Here, we fill this gap by studying how information spreads among 
a heterogeneous population of  ABPs that move in a two-dimensional space. 
Each particle or agent $i$  possesses a constant active speed  $v_i$ and the population is characterized by the distribution $p(v_i)$.  
Our goal is to obtain a theoretical understanding on spreading processes in heterogeneous active systems, 
without attempting to connect our study to any specific real-world system. 
To get insight into  the physics of such heterogeneous systems, we 
combine elements of kinetic theory~\cite{reif}, active matter~\cite{vicsek2012, marchetti2013}, and complex networks~\cite{newman2001random, callaway2000network, cohen2000resilience, cohen2001breakdown, mitra2008generalized, newman2002spread, castellano2010thresholds, boguna2003absence}. The used approach allows us to obtain analytical expressions for the spreading  threshold and demonstrate it depends on $p(v_i)$. 
Furthermore, we also investigate the effect of agent heterogeneity in transmission parameters or ``vaccination", i.e. we consider a sub-population of agents that do not participate in the spreading process, and show that there is an interplay between both heterogeneities, in activity and transmission parameters. 

It is important to stress that a system of moving agents cannot be reduced to a dynamical network where links among nodes are dynamically activated and inactivated~\cite{newman2002spread, castellano2010thresholds, boguna2003absence, mitra2008generalized}. Such a representation requires the existence of an underlying static network~\cite{newman2002spread,  boguna2003absence}.  
However, for a system of moving agents, it does not exist an unambiguous and unique way to determine such underlying network. 
To understand this, let us consider a system of $N$ (independent and persistent) random walkers (RW) moving in 2D. 
Whenever two RWs are at a distance less than $d_0$, we introduce a link between them. 
The recurrent property of RWs in 2D allows us to ensure that if we wait long enough, the resulting underlying network corresponds 
to a fully connected network, independently whether all the persistent RWs move at the same speed or exhibit distinct individuals speed, 
and independently of the speed distribution. 
On the other hand, if we consider a short time-interval, at low densities, the resulting network consists of 
mainly disconnected nodes. 
Despite this inherently difficulty associated to moving agent systems, here we explain how to characterize information spreading in active heterogeneous systems.


{\it Active system \& epidemic model.---}
We consider a system of $N$ heterogeneous ABPs moving on a  box of area $L\!\times\!L$ with periodic boundary conditions {\color{blue} at densities $\rho=N/L^2$ below the emergence of motility-induced phase separation (MIPS); in the presence of MIPS, the physics of the spreading dynamics is different~\cite{forgacs2022using}}. The equation of motion of the $i$-th particle is given by: 
\begin{eqnarray} 
\label{update_position} \dot{\mathbf{r}}_i &=& 
v_i \mathbf{e}[\theta_i]  -   \mu \sum_{j\neq i} 
{\mathbf{\nabla}_i U(\mathbf{r}_i, \mathbf{r}_j)} \, ,
\end{eqnarray} 
where $\mathbf{r}_i(t)$ refers to the position of the particle and $\theta_i(t)$ encodes (in 2D) the direction in which the self-propelling force is applied:   
$\mathbf{e}[\theta_i(t)] =
[\cos(\theta_i(t)),\sin(\theta_i(t))]$. 
We investigate two dynamics for $\theta_i(t)$ that lead qualitatively to identical results: (i) standard ABP, where $\dot{\theta}_i = \sqrt{2\lambda} \xi_i$, with 
$\xi_i$ a $\delta$-correlated noise, and (ii) classical  Run-\&-Tumble, where $\theta_i$ follows a Poisson process with rate $\lambda$, and new angles are randomly selected from the interval $[0, 2\pi)$. 
Particles interact via a short-range, repulsive potential $U$. We set $\mu=1$ in Eq.~(\ref{update_position}).  Identical results were obtained with WCA potential~\cite{weeks1971role} and   
$U(\textbf{r}_i , \textbf{r}_j)=a \left[ \left( \rvert \textbf{r}_i-\textbf{r}_j \rvert/2r\right)^{-1} -1\right]$, for $\rvert \textbf{r}_i-\textbf{r}_j \rvert < 2r$ and $0$ otherwise; see~\cite{peruani2019reaction}. The particle radius is set to $r=1$. 
Importantly, in Eq.~(\ref{update_position}),  $v_i$ denotes the $i$-th particle active speed, which is a unique, time-invariant feature of the particle. 
The active speed $v_i$ is drawn from probability density function (PDF) $p(v)$. 
\begin{figure*}[t]
    \centering
    \includegraphics[width=\textwidth]{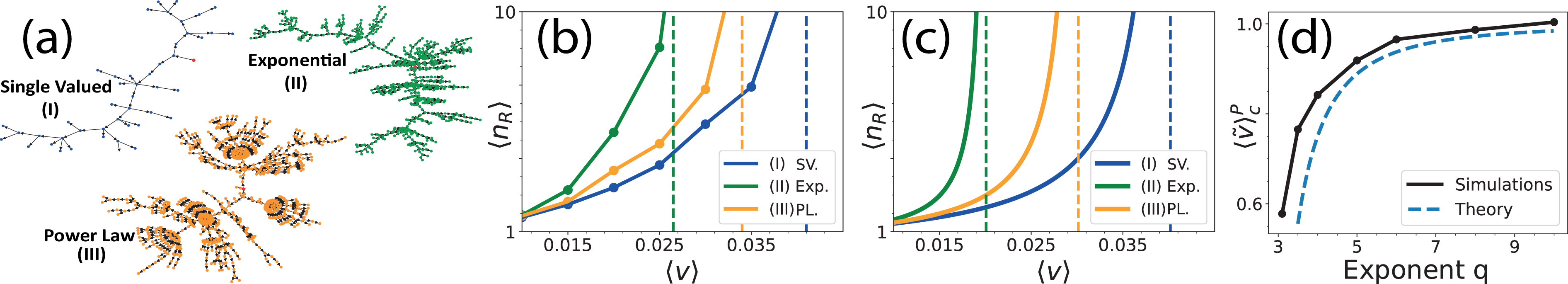}
    \caption{\label{fig:EpiThreshold} (a) Contagion networks for  three distinct active speed distribution $p(v)$ {\color{blue} that display exactly the same $\langle v \rangle=0.05$}.  Mean outbreak size $\langle n_R \rangle$  vs $\langle v \rangle$ in simulations (b) and  from  Eq.~(\ref{eq:n_R}) in (c). {\color{blue} Recall that in simulations $N$ is finite, while the theory assumes $N\to\infty$}. Vertical lines indicate epidemic thresholds; in (c)  correspond to Eq.~(\ref{eq:threshold_generic}).  (d) Epidemic threshold for the power-law distribution vs exponent $q$ in simulations and theory [Eq.~(\ref{eq:threshold_speed_Power})],  where  
    $ \langle \tilde v  \rangle^P_c = \langle v \rangle^P_c / \langle v \rangle^{SV}_c$. Parameters: $\rho\!=\!0.044$, $\beta\!=\!0.005$, $\lambda\!=\!0.01$,  $N\!=\!1024$, $P_{\text{inf}}\!=\!1$. $q\!=\!4$ in (b), (c), with averages  computed over $3000$ simulations. } 
\end{figure*}
We consider three different PDFs: 
\begin{equation}
\label{eq:distrib}
p(v) = 
  \begin{cases} 
   \delta(v-v_0) & \text{{\color{blue} single-valued}}  \\
    \exp[-v/ \bar{v}]/\bar{v}      &  \text{exponential} \\ 
    c\, v^{-q} & \text{power-law}
      \end{cases}
\end{equation}
where  $v_0$ and $\bar{v}$ are constants. The power-law distribution is non-zero  in the interval $S=[v_m, v_M]$, and $c$ is the normalization constant, and  $q\geq2$. 
{\color{blue} Note that the single-valued distribution corresponds to an homogenous system of identical active particles.}   
%

In addition, each particle possesses an internal (epidemic) state that adopts one of the following values: Susceptible (S), Infected (I), or Recovered (R). 
We consider the classical SIR dynamics that is summarized by the transitions: 
\begin{eqnarray}
\nonumber
\begin{array}{cc}
S+I \stackrel{\alpha}{\rightarrow}  2I , & 
I  \stackrel{\beta}{\rightarrow}       R \, ,
\end{array}
\end{eqnarray}
where $\alpha$ and $\beta$ are constant reaction rates. 
The reaction $S+I\to2I$ requires a particle in state $S$ and another one in state $I$  to keep physical contact 
for a finite time. 

{\it  Dynamics and epidemic threshold.---}
At $t=0$, we set all agents in state S, except for one that is set to state $I$, and let the system evolve. 
This initial infected agent eventually transmits the disease to $n$ other agents, which, in turn, pass over the disease 
to  other agents, and so on, until no agent in state $I$ remains in the system and the epidemic dynamics stops.  
The process can be represented by a contagion network as shown in Fig.~\ref{fig:EpiThreshold}(a); see also Fig. S1 in \cite{SM}. 
The quantity of interest is the size  $n_R$ of the contagion network, i.e. the number of agents in state R at $t \to \infty$, a.k.a. outbreak size. 
In Individual-Based Model (IBM) simulations, {\color{blue} where necessarily the system size $N$ is finite}, we observe that $\langle n_R\rangle$ -- where  $\langle \cdots \rangle$ denotes average over realizations -- 
increases with $\langle v \rangle$ --  where $\langle v^n \rangle = \int v^n p(v)\,dv$ --  for all active speed distributions $p(v)$ as shown in  Fig.~\ref{fig:EpiThreshold}(b). 
Each of these distributions exhibits a distinct critical point $\langle v \rangle_c$, i.e. vertical lines in Fig.~\ref{fig:EpiThreshold}(b), 
which implies that the threshold does not depend only on $\langle v \rangle$.  

In the following, we develop a theoretical argument, that neglects spatial correlations, but accounts for particle heterogeneity, to estimate the epidemic threshold.  
%
In kinetic theory of gases~\cite{reif}, the average number of collisions per time unit is approximately $v \sigma \rho$, where $\sigma \sim 4r$.    
%
%
This implies that if a particle is infected during a time $\beta^{-1}$, it will collide with $k = v \sigma \rho \beta^{-1}$ (provided $v  \lambda^{-1} > \rho^{-1/2}$). 
%
Evidently, each collision between a particle in state $S$ and another one in state $I$, does not necessarily result in the infection of the particle in state $S$. 
For a collision of duration $w$, the probability that the infection occurs is $1-\exp(-w\alpha)$~\cite{peruani2019reaction}. 
Averaging over all possible durations, we can compute the probability $P_{\text{inf}}$ that the infection occurs during a collision as 
$P_{\text{inf}}=\int^{\infty}_{0} p(w) [ 1-\exp(-w\alpha) ] dw $, where $p(w)$ is the PDF of observing a collision of duration $w$. 
In~\cite{peruani2008dynamics, peruani2019reaction}, it was found that $p(w)=\exp[-w/\langle w \rangle]/\langle w \rangle$, where $\langle w \rangle$ is the average collision duration. 
Inserting this expression into the definition of  $P_{\text{inf}}$, we obtain $P_{\text{inf}}=\frac{\alpha \langle w \rangle}{1+\alpha \langle w \rangle}$. 
Then, provided that an infected particle experiences $k$ collisions during a time $\beta^{-1}$, the probability that $\tilde{k}$ of those leads to an infection is $\binom{k}{\tilde{k}} (P_{\text{inf}})^{\tilde{k}} (1-P_{\text{inf}})^{k-\tilde{k}}$. 
Knowing the active speed PDF $p(v)$, we can estimate the probability $P(k)$ that a randomly chosen particle experiences $k$ collisions in a time $\beta^{-1}$ as $P(k) = p(\frac{k\beta}{\sigma \rho })\frac{\beta}{\sigma \rho}=p(v) \frac{1}{\phi}$, with $\phi = \sigma \rho \beta^{-1}$. 
%
Then, the probability $\tilde{p}(k; P_{\text{inf}})$ of transmitting the disease to $k$ other particles during a time $\beta^{-1}$ is: 
\begin{eqnarray}
\label{eq:prob_k}
\tilde{p}(k; P_{\text{inf}}) = \sum_{k'=k}^{\infty} {P}(k')  \binom{k'}{k} (P_{\text{inf}})^{k} (1-P_{\text{inf}})^{k'-k}
\end{eqnarray}
Using Eq.~(\ref{eq:prob_k}), we obtain the associated generating function $g_0(x; P_{\text{inf}}) = \sum_{k=0}^{\infty} \tilde{p}(k; P_{\text{inf}}) x^k = \sum_{k'=0}^{\infty} {P}(k') \left[1+(x-1) P_{\text{inf}} \right]^{k'}$ that can be expressed as: 
\begin{eqnarray}
\label{eq:generating_0}
 g_0(x; P_{\text{inf}}) = \int p(v)  \left[1+(x-1) P_{\text{inf}} \right]^{\phi v} \, dv \, .
\end{eqnarray}
%
%
%
From Eq.(\ref{eq:generating_0}) it is possible to compute $\langle k^n \rangle = \left[(x\, d/dx)^{n}g_0\right]\rvert_{x=1}$. 
%
Another quantity of interest is the probability that a given collision involves a particle that infects $k$ agents: 
$k\,\tilde{p}(k; P_{\text{inf}})/\langle k \rangle$ with $\langle k \rangle = g'_0(1; P_{\text{inf}})$, where $[\cdot]^{'}\equiv d[\cdot]/dx$. 
In particular, we are interested in the generating function associated to the probability that a random collision 
involves a contact with a particle that during a time $\beta^{-1}$ post-collision experiences $k-1$ ``successful" collisions,  
which can be expressed as $g_1(x; P_{\text{inf}})=g'_0(x; P_{\text{inf}})/g'_0(1; P_{\text{inf}})$. 
Using a fundamental result in percolation of complex networks~\cite{newman2001random, callaway2000network, cohen2000resilience, cohen2001breakdown, mitra2008generalized} that, based on $g_0$ and $g_1$ {\color{blue} and assuming an infinite system}, estimates the average cluster size as $1+g_0'/(1-g_1')|_{x=1}$, we obtain: 
%
%
\begin{eqnarray}
\label{eq:n_R}
\langle n_R \rangle = 1+ \frac{P_{\text{inf}} \phi \langle v \rangle}{1-P_{\text{inf}} \left[ \frac{\phi\langle v^2\rangle}{\langle v \rangle}-1 \right]} \, . 
\end{eqnarray}
%
%
Note that Eq.~(\ref{eq:n_R})  is valid only below the percolation threshold {\color{blue} and assumes $N\! \to \!\infty$}. 
%
The epidemic threshold takes the form: 
\begin{eqnarray}
\label{eq:threshold_generic}
P_{\text{inf}} = \frac{\alpha \langle \omega \rangle}{1+\alpha \langle \omega \rangle } = \frac{1}{\phi \frac{\langle v^2 \rangle}{\langle v \rangle} - 1} \, .
\end{eqnarray}
Assuming instantaneous transmission, i.e. $\alpha \to \infty$, $P_{\text{inf}}=1$ and Eq.(\ref{eq:threshold_generic}) reduces to: 
\begin{eqnarray}
\label{eq:threshold_speed}
\frac{\langle v^2 \rangle}{\langle v \rangle} = \frac{2}{\phi} \, .
\end{eqnarray}
Eqs.~(\ref{eq:threshold_generic}) and  (\ref{eq:threshold_speed}) put in evidence that the 
threshold depends not only on $\langle v \rangle$, but also on $\langle v^2 \rangle$. 
Note that the  average number of 
 secondary infections, $R_0$, is not an adequate control parameter since by definition $R_0 \sim \langle k \rangle \propto \langle v \rangle$. 
The obtained results indicate that different speed distributions -- even when they share the same $\langle v \rangle$ -- exhibit 
different epidemic thresholds; see Fig.~\ref{fig:EpiThreshold}(b) and (c). 
To illustrate this observation, let us use Eq.~(\ref{eq:threshold_speed}) to compute the critical average speed $\langle v \rangle^\mathcal{D}_c$ that ensures the possibility of giant  outbreaks for a distribution $p_{\mathcal{D}}(v)$. 
For {\color{blue} the single-valued distribution (SV)}, all agents exhibit the same speed $v_0$ -- thus $\langle v \rangle = v_0 $ -- and according to Eq.~(\ref{eq:threshold_speed}),  
  $\langle v \rangle^{SV}_{c} = 2/\phi$. 
For the exponential distribution (E), we obtain  $\langle v \rangle^{E}_c=1/\phi$. 
Finally, for the power-law (P) with $q=4$, we obtain  $\langle v \rangle^{E}_c < \langle v \rangle^{P}_c < \langle v \rangle^{SV}_c$. 
%
%
%
This proves that for the same density $\rho$, tumbling rate $\lambda$, and epidemic parameters, 
  $\langle v \rangle_c$ depends on the type of speed distribution, specifically on the ratio $\langle v^2 \rangle/\langle v \rangle$.  
{\color{blue} 
Fig.~\ref{fig:EpiThreshold} displays $\langle n_R \rangle$ in (finite size) IBM simulations, panel (b), and the theoretical estimate (that assumes $N\!\to\!\infty$)  given by Eq.~(\ref{eq:n_R}), panel (c); vertical lines correspond to Eq.~(\ref{eq:threshold_speed}). 
Simulations results, as the system size in increased, asymptotically approach the theoretical estimate, which does not involved any adjustable parameter; see Fig. S2 in \cite{SM} for a finite size comparison.}   
 
Let us focus on  the power-law speed distribution. For  $q>3$, 
 $\langle v^2 \rangle/\langle v \rangle = \langle v \rangle (q-2)^2/[(q-1)(q-3)]$ with $S=[v_m,\infty)$. 
Inserting this expression into Eq.~(\ref{eq:threshold_speed}), we obtain:
\begin{eqnarray}
\label{eq:threshold_speed_Power}
\langle v \rangle_c^{P}=  \frac{2}{\phi} \frac{(q-1)(q-3)}{(q-2)^2}\,.
\end{eqnarray}
From Eq.~(\ref{eq:threshold_speed_Power}), it is evident that $\lim_{ q\to3^{+}}  \langle v \rangle_c^{P} \to 0$, and thus 
the critical average speed required to observe  giant outbreaks vanishes; see Fig.~\ref{fig:EpiThreshold}(d). 
In other words, we expect giant outbreaks for any  set of parameters $\rho$, $\lambda$, $r$, $\alpha$ and $\beta$, 
for   $2 < q \leq 3$. 
In this range of $q$, infinite infection cascades take place for any epidemic parameter 
even when the vast majority of agents move very slowly: it is enough a small fraction of fast-moving agents to propagate the epidemics. 
{\color{blue} Fast moving agents become super-spreaders, since $v$ and $k$ are positively correlated.}  


\begin{figure*}[!]
    \centering
    \includegraphics[width=\textwidth]{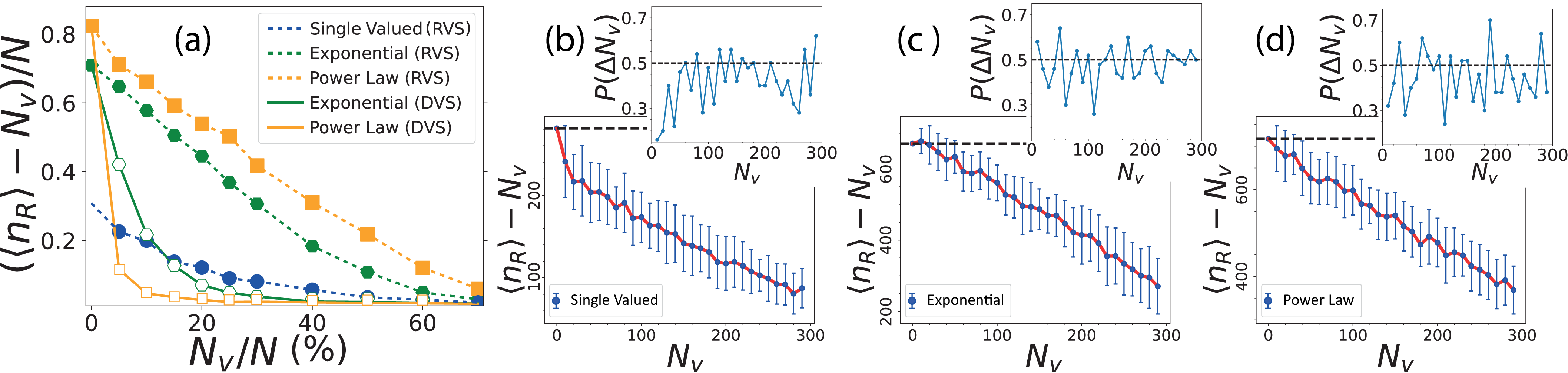} 
    \caption{\label{fig:vaccination}  (a) Normalized mean epidemic size, $[\langle n_R \rangle -N_v]/N$, average over $3000$ simulations, vs fraction of vaccinated population, $f=N_v/N$ for random (RVS) and directed (DVS) vaccination strategies. Mean epidemic size, $\langle n_R \rangle - N_v$ vs $N_v$ for the single-valued (b), exponential (c) and power-law (d)  distributions. Insets show  the probability that an increase of $\Delta N_v$  vaccinated agents leads to an increase in epidemic size. Parameters: as in Fig.~\ref{fig:EpiThreshold} and $\langle v \rangle = 0.05$; $S=[0.01, 4]$.}
\end{figure*}

{\it Vaccination strategies.---}
%
%
We have studied the impact of heterogeneity at the level of the active speed.  
However, at the level of epidemic parameters, the agent population was homogeneous. 
Let us consider that there exists also inter-individual variability in the agent's response to the disease, 
with a  sub-population of $N_v$ vaccinated agents -- implemented as $N_v$ agents set in state  R at $t=0$ -- which neither contract nor transmit the disease.  
The goal is to understand how the outbreak size, defined now as $\langle n_R \rangle - N_v$, depends on $N_v$. 
Next, we choose the vaccination strategy (VS) 
to distribute the $N_v$ vaccines among the agent population. 
We consider two strategies: random (RVS) and directed (DVS). 
The RVS consists in vaccinating randomly $N_v$ agents.   
In the DVS, on the other hand, we order agents according to their active speed, in decreasing order, and vaccinate the first $N_v$ in that list. 

We find that the RVS is not effective when applied to systems where agents exhibit heterogeneous active speeds, 
in comparison with the strong impact this strategy has on homogeneous agent populations; see Fig.~\ref{fig:vaccination}(a) dashed lines. 
The reason for this is that when we select randomly agents in the exponential and power-law distribution, it is likely 
that the chosen agents will possess a low active speed: slow agents are more abundant than fast agents. 
However, fast agents contribute significantly more in the spreading of the disease than slow agents, and thus decreasing the 
outbreak size requires  vaccinating a large fraction of agents.  
The RVS is an effective strategy only when all agents possess the same active speed. 
On the other hand, heterogeneous population of active agents are extremely sensitive to the DVS; see Fig.~\ref{fig:vaccination}(a) solid lines.  
This strategy ensures that vaccinated agents corresponds to those that otherwise will contribute the most to the spreading of the disease. 
The DVS leads to a dramatic reduction in the average epidemic size with the use of small number of vaccines [see also Fig. S3 in \cite{SM}]. 
%

{\it Vaccination as noise.---}
It is tempting to think that vaccinating $\Delta N_v$ extra individuals leads necessarily to a decrease in the outbreak size. 
However, we will see that this is not true. Adding $\Delta N_v$ extra vaccinated individuals (selected randomly) -- with $\Delta N_v \ll N$ --  can lead to either an increase or a decrease of the outbreak size for  all active speed distributions. 

To put in evidence this effect, we simulate the spatial dynamics of the agents and save 
the trajectory of every agent, including all collisions among agents. 
This allows us to replay the same spatial dynamics over and over. 
Using this spatial dynamics, we study 
the impact on outbreak size of vaccinating $\Delta N_v$ extra  agents.  
Two types of simulations are tested. 
One in which infected agent remains infected a duration that is obtained from a fixed list of random durations -- exponentially distributed with average $\beta^{-1}$ --  and another type of simulations where the infection duration is fixed to exactly $\beta^{-1}$.  
Note that the latter is a deterministic system. The probability $P(N_v, \Delta N_v)$ is computed as the fraction of simulations in which 
vaccinating additionally $\Delta N_v$ individuals to the previous $N_v$ vaccinated ones leads to an increase in outbreak size. 
%
We find that for any $N_v$, there exists always a probability that  adding  $\Delta N_v$ vaccinated agents leads to an increase of the outbreak size; 
see inset in (b), (c), (d) in Fig.~\ref{fig:vaccination}. 
However, this probability is  slightly smaller than $1/2$. 
This implies that though vaccinating  $\Delta N_v$  agents is more likely to lead to a decrease, an increase in the outbreak size cannot be excluded. 
Though in average over realizations, the outbreak size decreases with $N_v$; main panels in Fig.~\ref{fig:vaccination}(b)-(d). 
But, how is possible that  vaccinating $\Delta N_v$ extra agents leads to an increase in outbreak size? 
When we vaccinate $\Delta N_v$ extra agents, we also affect the time at which agents get initially infected. 
Vaccinating generates a new sequence of times at which agents contract the disease. Thus, the system explores  
different  ``infection paths" with different path lengths (outbreak sizes) [see also Fig. S4 and video in \cite{SM}].  


Beyond the obvious relevance of the obtained results within theoretical epidemiology,  
the obtained results represent a fundamental step towards a better understanding on how information propagates in heterogeneous active systems.  
For instance, the absence of an epidemic threshold for active speed distributions with fat enough tails suggests that 
the physics of systems with heterogeneous active agents is very different from the one of systems with identical agents. For instance, we speculate that strongly heterogeneous polar fluids may exhibit a unique order phase, and strongly  heterogeneous ABPs may not phase separate. 
Combining tools from kinetic theory and complex networks appears as a promising research direction that may pave the way to tackle several current challenging questions related to inter-particle variability within active matter theory.  
%


%
\begin{acknowledgments}

\paragraph{Acknowledgement}
BM and GS~acknowledges financial support from CONICET and SeCyT-UNC. 
F.P. acknowledge financial support from C.Y. Initiative of Excellence (grant Investissements d'Avenir ANR-16-IDEX-0008), INEX 2021 Ambition Project CollInt and Labex MME-DII, projects 2021-258 and 2021-297.
\end{acknowledgments}

\bibliographystyle{apsrev}


\begin{thebibliography}{42}
\expandafter\ifx\csname natexlab\endcsname\relax\def\natexlab#1{#1}\fi
\expandafter\ifx\csname bibnamefont\endcsname\relax
  \def\bibnamefont#1{#1}\fi
\expandafter\ifx\csname bibfnamefont\endcsname\relax
  \def\bibfnamefont#1{#1}\fi
\expandafter\ifx\csname citenamefont\endcsname\relax
  \def\citenamefont#1{#1}\fi
\expandafter\ifx\csname url\endcsname\relax
  \def\url#1{\texttt{#1}}\fi
\expandafter\ifx\csname urlprefix\endcsname\relax\def\urlprefix{URL }\fi
\providecommand{\bibinfo}[2]{#2}
\providecommand{\eprint}[2][]{\url{#2}}

\bibitem[{\citenamefont{Vicsek and Zafeiris}(2012)}]{vicsek2012}
\bibinfo{author}{\bibfnamefont{T.}~\bibnamefont{Vicsek}} \bibnamefont{and}
  \bibinfo{author}{\bibfnamefont{A.}~\bibnamefont{Zafeiris}},
  \bibinfo{journal}{Physics Reports} \textbf{\bibinfo{volume}{517}},
  \bibinfo{pages}{71} (\bibinfo{year}{2012}).

\bibitem[{\citenamefont{Marchetti et~al.}(2013)\citenamefont{Marchetti, Joanny,
  Ramaswamy, Liverpool, J.~Prost, and Simha}}]{marchetti2013}
\bibinfo{author}{\bibfnamefont{M.~C.} \bibnamefont{Marchetti}},
  \bibinfo{author}{\bibfnamefont{J.~F.} \bibnamefont{Joanny}},
  \bibinfo{author}{\bibfnamefont{S.}~\bibnamefont{Ramaswamy}},
  \bibinfo{author}{\bibfnamefont{T.~B.} \bibnamefont{Liverpool}},
  \bibinfo{author}{\bibfnamefont{M.~R.} \bibnamefont{J.~Prost}},
  \bibnamefont{and} \bibinfo{author}{\bibfnamefont{R.~A.} \bibnamefont{Simha}},
  \bibinfo{journal}{Rev. Mod. Phys.} \textbf{\bibinfo{volume}{85}},
  \bibinfo{pages}{1143} (\bibinfo{year}{2013}).

\bibitem[{\citenamefont{Frasca et~al.}(2008)\citenamefont{Frasca, Buscarino,
  Rizzo, Fortuna, and Boccaletti}}]{frasca08}
\bibinfo{author}{\bibfnamefont{M.}~\bibnamefont{Frasca}},
  \bibinfo{author}{\bibfnamefont{A.}~\bibnamefont{Buscarino}},
  \bibinfo{author}{\bibfnamefont{A.}~\bibnamefont{Rizzo}},
  \bibinfo{author}{\bibfnamefont{L.}~\bibnamefont{Fortuna}}, \bibnamefont{and}
  \bibinfo{author}{\bibfnamefont{S.}~\bibnamefont{Boccaletti}},
  \bibinfo{journal}{Phys. Rev. Lett.} \textbf{\bibinfo{volume}{100}},
  \bibinfo{pages}{044102} (\bibinfo{year}{2008}).

\bibitem[{\citenamefont{Riedel-Kruse et~al.}(2007)\citenamefont{Riedel-Kruse,
  M{\"{u}}ller, and Oates}}]{riedel07}
\bibinfo{author}{\bibfnamefont{I.~H.} \bibnamefont{Riedel-Kruse}},
  \bibinfo{author}{\bibfnamefont{C.}~\bibnamefont{M{\"{u}}ller}},
  \bibnamefont{and} \bibinfo{author}{\bibfnamefont{A.~C.} \bibnamefont{Oates}},
  \bibinfo{journal}{Science} \textbf{\bibinfo{volume}{317}},
  \bibinfo{pages}{1911} (\bibinfo{year}{2007}).

\bibitem[{\citenamefont{Peruani et~al.}(2010)\citenamefont{Peruani, Nicola, and
  Morelli}}]{peruani2010mobility}
\bibinfo{author}{\bibfnamefont{F.}~\bibnamefont{Peruani}},
  \bibinfo{author}{\bibfnamefont{E.~M.} \bibnamefont{Nicola}},
  \bibnamefont{and} \bibinfo{author}{\bibfnamefont{L.~G.}
  \bibnamefont{Morelli}}, \bibinfo{journal}{New Journal of Physics}
  \textbf{\bibinfo{volume}{12}}, \bibinfo{pages}{093029}
  (\bibinfo{year}{2010}).


\bibitem[{\citenamefont{Katz et~al.}(2011)\citenamefont{Katz, Tunstr{\o}m,
  Ioannou, Huepe, and Couzin}}]{katz2011inferring}
\bibinfo{author}{\bibfnamefont{Y.}~\bibnamefont{Katz}},
  \bibinfo{author}{\bibfnamefont{K.}~\bibnamefont{Tunstr{\o}m}},
  \bibinfo{author}{\bibfnamefont{C.~C.} \bibnamefont{Ioannou}},
  \bibinfo{author}{\bibfnamefont{C.}~\bibnamefont{Huepe}}, \bibnamefont{and}
  \bibinfo{author}{\bibfnamefont{I.~D.} \bibnamefont{Couzin}},
  \bibinfo{journal}{Proceedings of the National Academy of Sciences}
  \textbf{\bibinfo{volume}{108}}, \bibinfo{pages}{18720}
  (\bibinfo{year}{2011}).

\bibitem[{\citenamefont{Lopez et~al.}(2012)\citenamefont{Lopez, Gautrais,
  Couzin, and Theraulaz}}]{lopez2012behavioural}
\bibinfo{author}{\bibfnamefont{U.}~\bibnamefont{Lopez}},
  \bibinfo{author}{\bibfnamefont{J.}~\bibnamefont{Gautrais}},
  \bibinfo{author}{\bibfnamefont{I.~D.} \bibnamefont{Couzin}},
  \bibnamefont{and}
  \bibinfo{author}{\bibfnamefont{G.}~\bibnamefont{Theraulaz}},
  \bibinfo{journal}{Interface focus} \textbf{\bibinfo{volume}{2}},
  \bibinfo{pages}{693} (\bibinfo{year}{2012}).

\bibitem[{\citenamefont{G{\'o}mez-Nava
  et~al.}(2022)\citenamefont{G{\'o}mez-Nava, Bon, and
  Peruani}}]{gomez2022intermittent}
\bibinfo{author}{\bibfnamefont{L.}~\bibnamefont{G{\'o}mez-Nava}},
  \bibinfo{author}{\bibfnamefont{R.}~\bibnamefont{Bon}}, \bibnamefont{and}
  \bibinfo{author}{\bibfnamefont{F.}~\bibnamefont{Peruani}},
  \bibinfo{journal}{Nature Physics} \textbf{\bibinfo{volume}{18}},
  \bibinfo{pages}{1494} (\bibinfo{year}{2022}).

\bibitem[{\citenamefont{Bain and Bartolo}(2019)}]{bain2019dynamic}
\bibinfo{author}{\bibfnamefont{N.}~\bibnamefont{Bain}} \bibnamefont{and}
  \bibinfo{author}{\bibfnamefont{D.}~\bibnamefont{Bartolo}},
  \bibinfo{journal}{Science} \textbf{\bibinfo{volume}{363}},
  \bibinfo{pages}{46} (\bibinfo{year}{2019}).

\bibitem[{\citenamefont{Peruani and Sibona}(2008)}]{peruani2008dynamics}
\bibinfo{author}{\bibfnamefont{F.}~\bibnamefont{Peruani}} \bibnamefont{and}
  \bibinfo{author}{\bibfnamefont{G.~J.} \bibnamefont{Sibona}},
  \bibinfo{journal}{Physical review letters} \textbf{\bibinfo{volume}{100}},
  \bibinfo{pages}{168103} (\bibinfo{year}{2008}).

\bibitem[{\citenamefont{Peruani and Sibona}(2019)}]{peruani2019reaction}
\bibinfo{author}{\bibfnamefont{F.}~\bibnamefont{Peruani}} \bibnamefont{and}
  \bibinfo{author}{\bibfnamefont{G.~J.} \bibnamefont{Sibona}},
  \bibinfo{journal}{Soft matter} \textbf{\bibinfo{volume}{15}},
  \bibinfo{pages}{497} (\bibinfo{year}{2019}).

\bibitem[{\citenamefont{Norambuena et~al.}(2020)\citenamefont{Norambuena,
  Valencia, and Guzm{\'a}n-Lastra}}]{norambuena2020understanding}
\bibinfo{author}{\bibfnamefont{A.}~\bibnamefont{Norambuena}},
  \bibinfo{author}{\bibfnamefont{F.~J.} \bibnamefont{Valencia}},
  \bibnamefont{and}
  \bibinfo{author}{\bibfnamefont{F.}~\bibnamefont{Guzm{\'a}n-Lastra}},
  \bibinfo{journal}{Scientific Reports} \textbf{\bibinfo{volume}{10}},
  \bibinfo{pages}{20845} (\bibinfo{year}{2020}).

\bibitem[{\citenamefont{Rodr{\'\i}guez
  et~al.}(2022)\citenamefont{Rodr{\'\i}guez, Paoluzzi, Levis, and
  Starnini}}]{rodriguez2022epidemic}
\bibinfo{author}{\bibfnamefont{J.~P.} \bibnamefont{Rodr{\'\i}guez}},
  \bibinfo{author}{\bibfnamefont{M.}~\bibnamefont{Paoluzzi}},
  \bibinfo{author}{\bibfnamefont{D.}~\bibnamefont{Levis}}, \bibnamefont{and}
  \bibinfo{author}{\bibfnamefont{M.}~\bibnamefont{Starnini}},
  \bibinfo{journal}{Physical Review Research} \textbf{\bibinfo{volume}{4}},
  \bibinfo{pages}{043160} (\bibinfo{year}{2022}).

\bibitem[{\citenamefont{Gu et~al.}(2024)\citenamefont{Gu, Li, Gao, Su, Sun, and
  Wang}}]{gu2024influence}
\bibinfo{author}{\bibfnamefont{W.}~\bibnamefont{Gu}},
  \bibinfo{author}{\bibfnamefont{W.}~\bibnamefont{Li}},
  \bibinfo{author}{\bibfnamefont{F.}~\bibnamefont{Gao}},
  \bibinfo{author}{\bibfnamefont{S.}~\bibnamefont{Su}},
  \bibinfo{author}{\bibfnamefont{B.}~\bibnamefont{Sun}}, \bibnamefont{and}
  \bibinfo{author}{\bibfnamefont{W.}~\bibnamefont{Wang}},
  \bibinfo{journal}{Chaos: An Interdisciplinary Journal of Nonlinear Science}
  \textbf{\bibinfo{volume}{34}} (\bibinfo{year}{2024}).

\bibitem[{\citenamefont{Sajjadi et~al.}(2021)\citenamefont{Sajjadi, Hashemi,
  and Ghanbarnejad}}]{sajjadi2021social}
\bibinfo{author}{\bibfnamefont{S.}~\bibnamefont{Sajjadi}},
  \bibinfo{author}{\bibfnamefont{A.}~\bibnamefont{Hashemi}}, \bibnamefont{and}
  \bibinfo{author}{\bibfnamefont{F.}~\bibnamefont{Ghanbarnejad}},
  \bibinfo{journal}{Physical Review E} \textbf{\bibinfo{volume}{104}},
  \bibinfo{pages}{014313} (\bibinfo{year}{2021}).
  
 \bibitem[{\citenamefont{Te Vrugt et~al.}(2020)\citenamefont{Te Vrugt, Bickmann,
  and Wittkowski}}]{te2020effects}
\bibinfo{author}{\bibfnamefont{M.}~\bibnamefont{Te Vrugt}},
  \bibinfo{author}{\bibfnamefont{J.}~\bibnamefont{Bickmann}}, \bibnamefont{and}
  \bibinfo{author}{\bibfnamefont{R.}~\bibnamefont{Wittkowski}},
  \bibinfo{journal}{Nature Communications} \textbf{\bibinfo{volume}{11}},
  \bibinfo{pages}{5576} (\bibinfo{year}{2020}).
 

\bibitem[{\citenamefont{Forg{\'a}cs et~al.}(2022)\citenamefont{Forg{\'a}cs,
  Lib{\'a}l, Reichhardt, Hengartner, and Reichhardt}}]{forgacs2022using}
\bibinfo{author}{\bibfnamefont{P.}~\bibnamefont{Forg{\'a}cs}},
  \bibinfo{author}{\bibfnamefont{A.}~\bibnamefont{Lib{\'a}l}},
  \bibinfo{author}{\bibfnamefont{C.}~\bibnamefont{Reichhardt}},
  \bibinfo{author}{\bibfnamefont{N.}~\bibnamefont{Hengartner}},
  \bibnamefont{and}
  \bibinfo{author}{\bibfnamefont{C.}~\bibnamefont{Reichhardt}},
  \bibinfo{journal}{Scientific Reports} \textbf{\bibinfo{volume}{12}},
  \bibinfo{pages}{11229} (\bibinfo{year}{2022}).

\bibitem[{\citenamefont{Zhu et~al.}(2023)\citenamefont{Zhu, Shen, Dong, and
  Wang}}]{zhu2023spatial}
\bibinfo{author}{\bibfnamefont{Y.}~\bibnamefont{Zhu}},
  \bibinfo{author}{\bibfnamefont{R.}~\bibnamefont{Shen}},
  \bibinfo{author}{\bibfnamefont{H.}~\bibnamefont{Dong}}, \bibnamefont{and}
  \bibinfo{author}{\bibfnamefont{W.}~\bibnamefont{Wang}},
  \bibinfo{journal}{Plos one} \textbf{\bibinfo{volume}{18}},
  \bibinfo{pages}{e0286558} (\bibinfo{year}{2023}).

\bibitem[{\citenamefont{Rodr{\'\i}guez
  et~al.}(2019)\citenamefont{Rodr{\'\i}guez, Ghanbarnejad, and
  Egu{\'\i}luz}}]{rodriguez2019particle}
\bibinfo{author}{\bibfnamefont{J.~P.} \bibnamefont{Rodr{\'\i}guez}},
  \bibinfo{author}{\bibfnamefont{F.}~\bibnamefont{Ghanbarnejad}},
  \bibnamefont{and} \bibinfo{author}{\bibfnamefont{V.~M.}
  \bibnamefont{Egu{\'\i}luz}}, \bibinfo{journal}{Scientific reports}
  \textbf{\bibinfo{volume}{9}}, \bibinfo{pages}{6463} (\bibinfo{year}{2019}).

\bibitem[{\citenamefont{Zhao et~al.}(2022)\citenamefont{Zhao, Huepe, and
  Romanczuk}}]{zhao2022contagion}
\bibinfo{author}{\bibfnamefont{Y.}~\bibnamefont{Zhao}},
  \bibinfo{author}{\bibfnamefont{C.}~\bibnamefont{Huepe}}, \bibnamefont{and}
  \bibinfo{author}{\bibfnamefont{P.}~\bibnamefont{Romanczuk}},
  \bibinfo{journal}{Scientific reports} \textbf{\bibinfo{volume}{12}},
  \bibinfo{pages}{2588} (\bibinfo{year}{2022}).

\bibitem[{\citenamefont{Knebel et~al.}(2019)\citenamefont{Knebel, Ayali,
  Guershon, and Ariel}}]{knebel2019intra}
\bibinfo{author}{\bibfnamefont{D.}~\bibnamefont{Knebel}},
  \bibinfo{author}{\bibfnamefont{A.}~\bibnamefont{Ayali}},
  \bibinfo{author}{\bibfnamefont{M.}~\bibnamefont{Guershon}}, \bibnamefont{and}
  \bibinfo{author}{\bibfnamefont{G.}~\bibnamefont{Ariel}},
  \bibinfo{journal}{Science advances} \textbf{\bibinfo{volume}{5}},
  \bibinfo{pages}{eaav0695} (\bibinfo{year}{2019}).

\bibitem[{\citenamefont{Otte et~al.}(2021)\citenamefont{Otte, Ipi{\~n}a,
  Pontier-Bres, Czerucka, and Peruani}}]{otte2021statistics}
\bibinfo{author}{\bibfnamefont{S.}~\bibnamefont{Otte}},
  \bibinfo{author}{\bibfnamefont{E.~P.} \bibnamefont{Ipi{\~n}a}},
  \bibinfo{author}{\bibfnamefont{R.}~\bibnamefont{Pontier-Bres}},
  \bibinfo{author}{\bibfnamefont{D.}~\bibnamefont{Czerucka}}, \bibnamefont{and}
  \bibinfo{author}{\bibfnamefont{F.}~\bibnamefont{Peruani}},
  \bibinfo{journal}{Nature communications} \textbf{\bibinfo{volume}{12}},
  \bibinfo{pages}{1990} (\bibinfo{year}{2021}).

\bibitem[{\citenamefont{Gonzalez et~al.}(2008)\citenamefont{Gonzalez, Hidalgo,
  and Barabasi}}]{gonzalez2008understanding}
\bibinfo{author}{\bibfnamefont{M.~C.} \bibnamefont{Gonzalez}},
  \bibinfo{author}{\bibfnamefont{C.~A.} \bibnamefont{Hidalgo}},
  \bibnamefont{and} \bibinfo{author}{\bibfnamefont{A.-L.}
  \bibnamefont{Barabasi}}, \bibinfo{journal}{nature}
  \textbf{\bibinfo{volume}{453}}, \bibinfo{pages}{779} (\bibinfo{year}{2008}).

\bibitem[{\citenamefont{Paoluzzi et~al.}(2020)\citenamefont{Paoluzzi, Leoni,
  and Marchetti}}]{paoluzzi2020information}
\bibinfo{author}{\bibfnamefont{M.}~\bibnamefont{Paoluzzi}},
  \bibinfo{author}{\bibfnamefont{M.}~\bibnamefont{Leoni}}, \bibnamefont{and}
  \bibinfo{author}{\bibfnamefont{M.~C.} \bibnamefont{Marchetti}},
  \bibinfo{journal}{Soft Matter} \textbf{\bibinfo{volume}{16}},
  \bibinfo{pages}{6317} (\bibinfo{year}{2020}).

\bibitem[{\citenamefont{Levis et~al.}(2020)\citenamefont{Levis, Diaz-Guilera,
  Pagonabarraga, and Starnini}}]{levis2020flocking}
\bibinfo{author}{\bibfnamefont{D.}~\bibnamefont{Levis}},
  \bibinfo{author}{\bibfnamefont{A.}~\bibnamefont{Diaz-Guilera}},
  \bibinfo{author}{\bibfnamefont{I.}~\bibnamefont{Pagonabarraga}},
  \bibnamefont{and} \bibinfo{author}{\bibfnamefont{M.}~\bibnamefont{Starnini}},
  \bibinfo{journal}{Physical Review Research} \textbf{\bibinfo{volume}{2}},
  \bibinfo{pages}{032056} (\bibinfo{year}{2020}).

\bibitem[{\citenamefont{Paoluzzi et~al.}(2018)\citenamefont{Paoluzzi, Leoni,
  and Marchetti}}]{paoluzzi2018fractal}
\bibinfo{author}{\bibfnamefont{M.}~\bibnamefont{Paoluzzi}},
  \bibinfo{author}{\bibfnamefont{M.}~\bibnamefont{Leoni}}, \bibnamefont{and}
  \bibinfo{author}{\bibfnamefont{M.~C.} \bibnamefont{Marchetti}},
  \bibinfo{journal}{Physical Review E} \textbf{\bibinfo{volume}{98}},
  \bibinfo{pages}{052603} (\bibinfo{year}{2018}).

\bibitem[{\citenamefont{Gascuel et~al.}(2023)\citenamefont{Gascuel, Rahmani,
  Bon, and Peruani}}]{gascuel2023generic}
\bibinfo{author}{\bibfnamefont{H.-M.} \bibnamefont{Gascuel}},
  \bibinfo{author}{\bibfnamefont{P.}~\bibnamefont{Rahmani}},
  \bibinfo{author}{\bibfnamefont{R.}~\bibnamefont{Bon}}, \bibnamefont{and}
  \bibinfo{author}{\bibfnamefont{F.}~\bibnamefont{Peruani}},
  \bibinfo{journal}{arXiv preprint arXiv:2311.06208}  (\bibinfo{year}{2023}).

\bibitem[{\citenamefont{Huang et~al.}(2016)\citenamefont{Huang, Ding, Feng, and
  Pan}}]{huang2016epidemic}
\bibinfo{author}{\bibfnamefont{Y.}~\bibnamefont{Huang}},
  \bibinfo{author}{\bibfnamefont{L.}~\bibnamefont{Ding}},
  \bibinfo{author}{\bibfnamefont{Y.}~\bibnamefont{Feng}}, \bibnamefont{and}
  \bibinfo{author}{\bibfnamefont{J.}~\bibnamefont{Pan}},
  \bibinfo{journal}{Journal of Statistical Mechanics: Theory and Experiment}
  \textbf{\bibinfo{volume}{2016}}, \bibinfo{pages}{103501}
  (\bibinfo{year}{2016}).

\bibitem[{\citenamefont{Marchetti et~al.}(2016)\citenamefont{Marchetti, Fily,
  Henkes, Patch, and Yllanes}}]{marchetti2016minimal}
\bibinfo{author}{\bibfnamefont{M.~C.} \bibnamefont{Marchetti}},
  \bibinfo{author}{\bibfnamefont{Y.}~\bibnamefont{Fily}},
  \bibinfo{author}{\bibfnamefont{S.}~\bibnamefont{Henkes}},
  \bibinfo{author}{\bibfnamefont{A.}~\bibnamefont{Patch}}, \bibnamefont{and}
  \bibinfo{author}{\bibfnamefont{D.}~\bibnamefont{Yllanes}},
  \bibinfo{journal}{Current Opinion in Colloid \& Interface Science}
  \textbf{\bibinfo{volume}{21}}, \bibinfo{pages}{34} (\bibinfo{year}{2016}).

\bibitem[{\citenamefont{Kumar et~al.}(2021)\citenamefont{Kumar, Singh, Giri,
  and Mishra}}]{kumar2021effect}
\bibinfo{author}{\bibfnamefont{S.}~\bibnamefont{Kumar}},
  \bibinfo{author}{\bibfnamefont{J.~P.} \bibnamefont{Singh}},
  \bibinfo{author}{\bibfnamefont{D.}~\bibnamefont{Giri}}, \bibnamefont{and}
  \bibinfo{author}{\bibfnamefont{S.}~\bibnamefont{Mishra}},
  \bibinfo{journal}{Physical Review E} \textbf{\bibinfo{volume}{104}},
  \bibinfo{pages}{024601} (\bibinfo{year}{2021}).

\bibitem[{\citenamefont{Paoluzzi et~al.}(2022)\citenamefont{Paoluzzi, Levis,
  and Pagonabarraga}}]{paoluzzi2022motility}
\bibinfo{author}{\bibfnamefont{M.}~\bibnamefont{Paoluzzi}},
  \bibinfo{author}{\bibfnamefont{D.}~\bibnamefont{Levis}}, \bibnamefont{and}
  \bibinfo{author}{\bibfnamefont{I.}~\bibnamefont{Pagonabarraga}},
  \bibinfo{journal}{Communications Physics} \textbf{\bibinfo{volume}{5}},
  \bibinfo{pages}{111} (\bibinfo{year}{2022}).

\bibitem[{\citenamefont{Miguel et~al.}(2018)\citenamefont{Miguel, Parley, and
  Pastor-Satorras}}]{miguel2018effects}
\bibinfo{author}{\bibfnamefont{M.~C.} \bibnamefont{Miguel}},
  \bibinfo{author}{\bibfnamefont{J.~T.} \bibnamefont{Parley}},
  \bibnamefont{and}
  \bibinfo{author}{\bibfnamefont{R.}~\bibnamefont{Pastor-Satorras}},
  \bibinfo{journal}{Physical review letters} \textbf{\bibinfo{volume}{120}},
  \bibinfo{pages}{068303} (\bibinfo{year}{2018}).

\bibitem[{\citenamefont{Reif}(1965)}]{reif}
\bibinfo{author}{\bibfnamefont{F.}~\bibnamefont{Reif}},
  \emph{\bibinfo{title}{Fundamentals of statistical and thermal physics}}
  (\bibinfo{publisher}{McGraw-Hill, Singapore}, \bibinfo{year}{1965}).

\bibitem{SM} See Supplemental Material at http://xxx.xxx/
for further supplemental figures and movies. 

\bibitem[{\citenamefont{Newman et~al.}(2001)\citenamefont{Newman, Strogatz, and
  Watts}}]{newman2001random}
\bibinfo{author}{\bibfnamefont{M.~E.} \bibnamefont{Newman}},
  \bibinfo{author}{\bibfnamefont{S.~H.} \bibnamefont{Strogatz}},
  \bibnamefont{and} \bibinfo{author}{\bibfnamefont{D.~J.} \bibnamefont{Watts}},
  \bibinfo{journal}{Physical review E} \textbf{\bibinfo{volume}{64}},
  \bibinfo{pages}{026118} (\bibinfo{year}{2001}).

\bibitem[{\citenamefont{Callaway et~al.}(2000)\citenamefont{Callaway, Newman,
  Strogatz, and Watts}}]{callaway2000network}
\bibinfo{author}{\bibfnamefont{D.~S.} \bibnamefont{Callaway}},
  \bibinfo{author}{\bibfnamefont{M.~E.} \bibnamefont{Newman}},
  \bibinfo{author}{\bibfnamefont{S.~H.} \bibnamefont{Strogatz}},
  \bibnamefont{and} \bibinfo{author}{\bibfnamefont{D.~J.} \bibnamefont{Watts}},
  \bibinfo{journal}{Physical review letters} \textbf{\bibinfo{volume}{85}},
  \bibinfo{pages}{5468} (\bibinfo{year}{2000}).

\bibitem[{\citenamefont{Cohen et~al.}(2000)\citenamefont{Cohen, Erez,
  Ben-Avraham, and Havlin}}]{cohen2000resilience}
\bibinfo{author}{\bibfnamefont{R.}~\bibnamefont{Cohen}},
  \bibinfo{author}{\bibfnamefont{K.}~\bibnamefont{Erez}},
  \bibinfo{author}{\bibfnamefont{D.}~\bibnamefont{Ben-Avraham}},
  \bibnamefont{and} \bibinfo{author}{\bibfnamefont{S.}~\bibnamefont{Havlin}},
  \bibinfo{journal}{Physical review letters} \textbf{\bibinfo{volume}{85}},
  \bibinfo{pages}{4626} (\bibinfo{year}{2000}).

\bibitem[{\citenamefont{Cohen et~al.}(2001)\citenamefont{Cohen, Erez,
  Ben-Avraham, and Havlin}}]{cohen2001breakdown}
\bibinfo{author}{\bibfnamefont{R.}~\bibnamefont{Cohen}},
  \bibinfo{author}{\bibfnamefont{K.}~\bibnamefont{Erez}},
  \bibinfo{author}{\bibfnamefont{D.}~\bibnamefont{Ben-Avraham}},
  \bibnamefont{and} \bibinfo{author}{\bibfnamefont{S.}~\bibnamefont{Havlin}},
  \bibinfo{journal}{Physical review letters} \textbf{\bibinfo{volume}{86}},
  \bibinfo{pages}{3682} (\bibinfo{year}{2001}).

\bibitem[{\citenamefont{Mitra et~al.}(2008)\citenamefont{Mitra, Ganguly, Ghose,
  and Peruani}}]{mitra2008generalized}
\bibinfo{author}{\bibfnamefont{B.}~\bibnamefont{Mitra}},
  \bibinfo{author}{\bibfnamefont{N.}~\bibnamefont{Ganguly}},
  \bibinfo{author}{\bibfnamefont{S.}~\bibnamefont{Ghose}}, \bibnamefont{and}
  \bibinfo{author}{\bibfnamefont{F.}~\bibnamefont{Peruani}},
  \bibinfo{journal}{Physical Review E} \textbf{\bibinfo{volume}{78}},
  \bibinfo{pages}{026115} (\bibinfo{year}{2008}).

\bibitem[{\citenamefont{Newman}(2002)}]{newman2002spread}
\bibinfo{author}{\bibfnamefont{M.~E.} \bibnamefont{Newman}},
  \bibinfo{journal}{Physical review E} \textbf{\bibinfo{volume}{66}},
  \bibinfo{pages}{016128} (\bibinfo{year}{2002}).

\bibitem[{\citenamefont{Castellano and
  Pastor-Satorras}(2010)}]{castellano2010thresholds}
\bibinfo{author}{\bibfnamefont{C.}~\bibnamefont{Castellano}} \bibnamefont{and}
  \bibinfo{author}{\bibfnamefont{R.}~\bibnamefont{Pastor-Satorras}},
  \bibinfo{journal}{Physical review letters} \textbf{\bibinfo{volume}{105}},
  \bibinfo{pages}{218701} (\bibinfo{year}{2010}).

\bibitem[{\citenamefont{Bogun{\'a} et~al.}(2003)\citenamefont{Bogun{\'a},
  Pastor-Satorras, and Vespignani}}]{boguna2003absence}
\bibinfo{author}{\bibfnamefont{M.}~\bibnamefont{Bogun{\'a}}},
  \bibinfo{author}{\bibfnamefont{R.}~\bibnamefont{Pastor-Satorras}},
  \bibnamefont{and}
  \bibinfo{author}{\bibfnamefont{A.}~\bibnamefont{Vespignani}},
  \bibinfo{journal}{Physical review letters} \textbf{\bibinfo{volume}{90}},
  \bibinfo{pages}{028701} (\bibinfo{year}{2003}).

\bibitem[{\citenamefont{Weeks et~al.}(1971)\citenamefont{Weeks, Chandler, and
  Andersen}}]{weeks1971role}
\bibinfo{author}{\bibfnamefont{J.~D.} \bibnamefont{Weeks}},
  \bibinfo{author}{\bibfnamefont{D.}~\bibnamefont{Chandler}}, \bibnamefont{and}
  \bibinfo{author}{\bibfnamefont{H.~C.} \bibnamefont{Andersen}},
  \bibinfo{journal}{The Journal of chemical physics}
  \textbf{\bibinfo{volume}{54}}, \bibinfo{pages}{5237} (\bibinfo{year}{1971}).

\end{thebibliography}

\end{document}